\documentclass[
]{elsart}
\usepackage{graphicx}
\journal{Physics Letters B}
\def\bfig{\begin{figure}}
\def\efig{\end{figure}}
\def\bc{\begin{center}}
\def\ec{\end{center}}
\def\1{\'{\i}}
\newcommand{\R}{\mathop{\mathrm{Re}}}
\newcommand{\be}{\begin{equation}}
\newcommand{\ee}{\end{equation}}
\newcommand{\bea}{\begin{eqnarray}}
\newcommand{\beas}{\begin{eqnarray*}}
\newcommand{\eea}{\end{eqnarray}}
\newcommand{\eeas}{\end{eqnarray*}}
\newcommand{\ba}{\begin{array}}
\newcommand{\ea}{\end{array}}
\newcommand{\bt}{\begin{table}}
\newcommand{\et}{\end{table}}
\newcommand{\btr}{\begin{tabular}}
\newcommand{\etr}{\end{tabular}}
\newcommand{\bi}{\begin{itemize}}
\newcommand{\ei}{\end{itemize}}
\newcommand{\ben}{\begin{enumerate}}
\newcommand{\een}{\end{enumerate}}

\begin{document}
\begin{frontmatter}
\title{Probing the mechanism of EWSB with a Rho parameter defined in terms
of Higgs couplings}

\author{J.~L.~D\'\i az-Cruz} \and
\ead{ldiaz@sirio.ifuap.buap.mx}
\author{D.~A.~L\'opez-Falc\'on\corauthref{cor}}
\ead{dennys@sirio.ifuap.buap.mx},
\ead[url]{www.ifuap.buap.mx/$\sim$dennys/}
\corauth[cor]{Corresponding author.}
\address{Instituto de F\'\i sica, BUAP, Apartado Postal J-48, Puebla,
Pue. 72570 M\'exico}
\begin{abstract}
A definition of the rho parameter based on the Higgs couplings
with the gauge bosons, $\rho_{H_i}\equiv\frac{g_{H_iWW}}{g_{H_iZZ}c^2_W}$, 
is proposed as a new probe into the origin of the mechanism of electroweak
symmetry breaking. While $\rho_{h_{SM}}=1$ holds in the standard model,
deviations from one for $\rho_{H_i}$ are predicted in models with extended
Higgs sector. We derive a general expresion of $\rho_{H_i}$ for a model
with arbitrary Higgs multiplets, and discuss its size within the context
of specific models with Higgs triplets, including the ``Little Higgs''
models recently proposed. We find the even for Higgs sectors that
incorporate the custodial symmetry to make $\rho=1$, one could have
$\rho_{H_i}\neq1$, which could be tested at the level of a few percent,
with the precision Higgs meassurements expected at the next linear
collider (NLC).
\end{abstract}
\end{frontmatter}

\section{Introduction}	\label{sec:intro}

The mechanism of electroweak symmetry breaking (EWSB), which is
triggered spontaneously through a Higgs doublet in the minimal
standard model (SM), has remained without direct experimental
verification so far. Precision  measurements of electroweak 
observables constrain the Higgs mass below about 200 GeV at  95\%
CL \cite{Langacker:2002sy,Villa:2002zt,Chanowitz:2002cd} within
the standard model. Thus, it is expected that a Higgs particle
could be discovered at the Run 2 of the Tevatron, provided
sufficient luminosity is achieved~\cite{Carena:2000yx}. But it is
intriguing to notice that the EW observables prefer a SM
like  Higgs with mass below 114.1 GeV
\cite{Langacker:2002sy,Chanowitz:2002cd}, which is the present
lower limit from LEP 2. The data indicate that the Higgs boson
should have already been discovered \cite{Langacker:2002sy}, and
the fact that it was not, could be taken as a hint of new physics,
which could be related with the freedom to choose the Higgs 
sector \cite{Chanowitz:2002cd}. Extensions of the Higgs sector have been
proposed for a while \cite{hhunter}, and in particular models with Higgs
triplets (real or complex) have been considered well motivated, partly
because such representations arise in the context of left-right symmetric
models \cite{LRmodels}, or are associated with low-energy mechanisms aimed
to generate neutrino masses \cite{lownum}, as an alternative to the usual
see-saw mechanism. More recently, Higgs triplets with $\mathcal{O}$
(TeV) masses, have been predicted in connection with the so-called
``Little Higgs'' models \cite{littleH}, which attempt to explain the
required lightness of the Higgs as being associated with a global
symmetry.

Models with Higgs triplets can violate the custodial
symmetry SU$(2)_c$ of the Higgs-Gauge sectors. This symmetry
protects the relation between the gauge boson masses and the weak
mixing angle, which can be conveniently parameterized through
Veltman's rho parameter~\cite{rhoVeltman}, i.e. $\rho=m^2_W/m^2_Zc^2_W$, 
which is equal to one at tree-level; loop corrections to this parameter
could be very important, as it was exemplified by the prediction of a top
quark heavier than originally expected. However, when one considers
models with Higgs triplets, with their neutral component acquiring a
v.e.v. that contributes to EWSB, then the $\rho$ parameter could deviate
from one even at tree-level. Several Higgs triplets, with ad hoc quantum
numbers, are required in order to preserve the custodial 
symmetry~\cite{georgitrip}.

A simple analysis of the SM Lagrangian reveals that the Gauge boson masses
and their Higgs couplings originate from the terms:
\be
(D^\mu \Phi)^\dagger(D_\mu \Phi)= \Phi^{0*} \Phi^0
            [ g^2 W^{+\mu} W^{-}_{\mu}+ {g'}^2 Z^{\mu} Z_{\mu}]+\cdots
\ee

After SSB, one can write the neutral component in terms of the SM Higgs
boson ($h_0$) and the Goldstone boson ($G_Z^0$), i.e. 
$\Phi^0=(v+h^0+iG^0_Z)/\sqrt{2}$, and the gauge bosons ($W^\pm$
and $Z^0$) acquire the masses: $m^2_W=g^2v^2/4$ and $m^2_Z={g'}^2v^2/4$, 
respectively, with $g'=g/c_W$. In this case it happens that the same
source of EWSB that contributes to the Gauge boson masses, induces the
Higgs-Gauge couplings, which in turn are given by:
$g_{hWW}=g^2v/2$, $g_{hZZ}={g'}^2v/2$, and therefore one can define the
parameters $\rho$ and $\rho_h$, which satisfy:
\be
\rho\equiv\frac{m^2_W}{m^2_Zc^2_W} = 1 =
\frac{g_{hWW}}{g_{hZZ}c^2_W}\equiv\rho_h.
\ee

We could also express this result by saying that both the neutral 
Higgs and Goldstone boson have the same couplings to the gauge bosons in
the SM. Thus, if the SM is the correct theory of EWSB, a measurement of
the Higgs-Gauge couplings should give $\rho_h=1$. However, small deviation
from one for $\rho_H$ can be expected to appear because of radiative
effects, while the experimental value of $\rho_h$ will deviate from one
because of the systematic and statistical errors.

Furthermore, when one considers physics beyond the SM aimed to explain
EWSB, it is conceivable that the Goldstone bosons could have a different
origin from other neutral scalar of the model, as it could happen in
composite scenarios. Alternatively, even if both the Higgs and Goldstone
bosons have a common origin, their Higgs-Gauge boson couplings could have
different values, either because of mixing factors or because of 
renormalization effects. In all these cases, one would have
$\rho \neq \rho_H$. Given the possibility that an scalar particle could be
detected in the near future, it will be important to verify whether this
particle is indeed a type of Higgs boson, and the parameter $\rho_H$ could
play a major role in this regard. This will be illustrated in the next
sections with several examples.

The organization of this letter goes as follows: In
section~\ref{sec:multiplets}, we shall present a general expression for
$\rho_H$ for a Higgs multiplet of arbitrary isospin $T$ and hypercharge
$Y$; its size is discussed in detail within the context of a minimal
extension of the SM that includes one doublet and a real $(Y=0)$ Higgs
triplet; one of our main result is the argument that $\rho\simeq1$ does
not implies $\rho_H=1$. We shall also evaluate a similar parameter, but in
terms of the Higgs decay widths, which would be closer to the output from
future high-precision experiments for the Higgs boson. We then discuss, in
section~\ref{sec:modeling}, a model with extended Higgs sectors, which do
respects the custodial symmetry, i.e. $\rho=1$, but the Higgs particles do
not necessarily satisfy $\rho_{H_i}=1$. Then, in
section~\ref{sec:littlehiggs} we shall discuss the above parameter, for
the Higgs sector that arises within the context of the ``Little Higgs''
model. Finally, we shall present our conclusions in
section~\ref{sec:conclusion}.

\section{Higgs multiplets and the Rho parameter}
\label{sec:multiplets}

\subsection{A general expression for $\rho_H$}
\label{general}

Let us consider a model with an arbitrary Higgs sector,
consisting of a number of Higgs multiplets $\Phi_K$ of
isospin $T_K$ and hypercharge $Y_K$. From the expression for
the kinetic terms, written in terms of the covariant
derivative, one obtains the gauge boson masses, which
satisfy the following expression for the rho parameter,
\be
\rho=\frac{\sum_K [T_K (T_K+1) -\frac{1}{4} Y^2_K]v^2_K c_K}
      {\sum_K \frac{1}{2} Y^2_K v^2_K }
\ee
where $v_K$ denotes the v.e.v. of the neutral component of the
Higgs multiplet, while $c_K=1/2\,(1)$ for real (complex) representations.
It is well known that Higgs representations for which
$T_K (T_K+1)=\frac{3}{4} Y^2_K$, satisfy $\rho=1$, regardless of their
v.e.v.'s. Examples of this case are: $(T,Y)=(1/2,1), (3,4),\ldots$.
Alternatively, one could choose ad hoc v.e.v.'s for models with several 
types of Higgs multiplets, such as triplets, to have $\rho=1$.

On the other hand, when one writes down the Gauge boson coupling with the
neutral Higgs components $\Phi_K^0$, which are weak eigenstates, $\rho_H$
satisfies a similar relation, namely:
\be
\rho_{\phi^0_K}=\frac{[T_K (T_K+1)-\frac{1}{4} Y^2_K]v^2_K c_K}
      {\frac{1}{2} Y^2_K v^2_K }
\ee
Thus, whatever choice makes $\rho=1$ for the Higgs multiplet
$\Phi_K$, it will also make $\rho_{\phi^0_K}=1$. However, when
one has several multiplets, one needs to consider the Higgs
mass eigenstates instead, which are indeed the ones that could
be detected and probed at future colliders. Thus, we have to
consider the rotations that diagonaliazes the real parts of
the neutral components, such that the Higgs mass eigenstates
$H_i$ are related to the weak eigenstates $\R \phi^0_K$
as: $\R \phi^0_K= U_{Ki} H_i$. Then, the rho parameter
for the Higgs bosons $H_i$ is given by:
\be	\label{rhoHgen}
\rho_{H_i}=\frac{\sum_K [T_K (T_K+1)-\frac{1}{4} Y^2_K]v^2_K c_K U_{Ki}}
      {\sum_K \frac{1}{2} Y^2_K v^2_K U_{Ki}}
\ee
From this important relation, we can discuss several consequences:
\ben
\item For models that contains several Higgs multiplets of the same type
(say doublets), for which $T_K(T_K+1)=\frac{1}{4}Y_K^2$, one gets
$\rho_{H_i}=1$ (as well as $\rho=1$), because $U_{Ki}$ factorize out in
Ec.~(\ref{rhoHgen}).
\item On the other hand for a model that includes doublets and some other
multiplet (say triplets), for which $\rho\simeq1$ is satisfied with a
hierarchy of v.e.v.'s, i.e. $v_K\ll v_D$, then one has that $\rho_{H_i}$
could be significantly different from one (as will be shown next).
\item Finally, if one makes $\rho=1$ by arranging the v.e.v.'s of several
multiplets (as in the model to be discussed in section 3), then because
of the factors $U_{Ki}$, it turns out that in general $\rho=1$ does not
necessarily imply $\rho_H=1$, and this could provide an important test of
the type of Higgs multiplet that participates in EWSB. 
\een

\subsection{A model with one doublet and one real triplet}
\label{sec:modeling}

We shall evaluate now the size of $\rho_H$ for an extension of the SM,
where the Higgs sector includes one real $(Y=0)$ Higgs triplet, 
$\Xi=(\xi^+,\xi^0,\xi^-)$, in addition to the usual SM Higgs doublet
$\Phi$. The Higgs potential of the model is written as \cite{1D_11T_0}:
\bea V(\Phi,\Xi)&=&-\mu_d^2\Phi^\dag\Phi+\lambda_1(\Phi^\dag\Phi)^2
        -\mu_{tr}^2\Xi^\dag\Xi+\lambda_2(\Xi^\dag\Xi)^2\nonumber\\
       && +\lambda_3\Phi^\dag\Phi\Xi^\dag\Xi
        -\mu_{dtr}[\Phi^\dag(\Xi_{lin}\cdot\tau)\Phi],
\eea 
where the last term involves the linear form, namely:
\be
\Xi_{lin}=(\frac{1}{\sqrt{2}}(\xi^++\xi^-),\frac{i}{\sqrt{2}}(\xi^+-\xi^-),\xi^0)
\ee
and $\tau$ is the vector of Pauli's spin matrices.

After constructing the mass matrices, and performing its diagonalization,
we arrive to the following mass eigenstates:
\be
    \left(\ba{c}H^0\\h^0\ea\right)=
        \left(\ba{cc}
        \cos{\alpha}&\sin{\alpha}\\
        -\sin{\alpha}&\cos{\alpha}
        \ea\right)
            \left(\ba{c}h_d\\h_{tr}\ea\right),   \label{rotalpha1d1tr}
\ee where $h_d=\R \Phi^0$ and $h_{tr}=\R \xi^0$; while the mixing angle,
$\alpha$, is defined by: \be
    \tan{2\alpha}=\frac{4v_Dv_{T0}(2\lambda_3v_{T0}+\sqrt{2}\mu_{dtr})}
{(8\lambda_1v_{T0}+\sqrt{2}\mu_{dtr})v_D^2-8\lambda_2v_{T0}^3},
    \label{alpha1d1tr}
\ee
here $v_D=\left<h_d\right>$ and $v_{T0}=\left<h_{tr}\right>$.

The Higgs-Gauge Lagrangian is given by: \be 
{\mathcal{T}}_{\mathrm{Cin}}=(D^\mu_\Phi\Phi)^\dag(D_{\Phi_\mu}\Phi)+(D^\mu_\Xi\Xi)^\dag(D_{\Xi_\mu}\Xi),\ee
where
\be D^\mu_\Phi=\left(\ba{cc}
\partial^\mu+igs_W(A^\mu-Z^\mu t_W)&\frac{i}{\sqrt{2}}gW^{\mu+}\\
\frac{i}{\sqrt{2}}gW^{\mu-}&\partial^\mu-\frac{i}{2}\frac{g}{c_W}Z^\mu
        \ea\right),	\label{DcovD_1fis}
\ee
and
\be
D_\Xi^\mu=\left(\ba{ccc}
\partial^\mu+ig(Z^\mu c_W+A^\mu s_W)&igW^{\mu+}&0\\
        igW^{\mu-}&\partial^\mu&igW^{\mu+}\\
        0&igW^{\mu-}&\partial^\mu-ig(Z^\mu c_W+A^\mu s_W)
        \ea\right).  \label{DcovT_0fis}
\ee

From this Lagrangian we can identify the masses of the Gauge bosons:
\bea 
m_W^2&=&\frac{g^2}{4}(v_D^2+4v_{T0}^2),\\
m_Z^2&=&\frac{m_W^2}{c^2_W}\left(\frac{v_D^2}{v_D^2+4v_{T0}^2}\right),\nonumber
\eea
and the couplings $hWW$, $hZZ$, $HWW$ and $HZZ$, where $h$ ($H$) 
corresponds to the lighter SM-like (heavier) neutral Higgs mass
eigenstate:
\bea
g_{hWW}&=&gm_W\cos\alpha\left[1 +\tan\alpha
                    \left(\frac{\Delta
\rho}{\rho}\right)^{1/2}\right],\nonumber\\
g_{hZZ}&=&\frac{gm_W}{c^2_W}\cos\alpha,\nonumber\\
g_{HWW}&=&gm_W\sin\alpha\left[\cot\alpha
                    \left(\frac{\Delta
\rho}{\rho}\right)^{1/2}-1\right],\nonumber\\
g_{HZZ}&=&\frac{gm_W}{c^2_W}\sin\alpha.
\eea

Therefore in this model we
have: $\rho=1+\frac{4v_{T0}^2}{v_D^2}\equiv1+\tan^2\beta$ and
$\rho_h^2=[1 +\tan\alpha (\frac{\Delta \rho}{\rho})^{1/2}]^2$ and 
$\rho_H^2=[\cot\alpha (\frac{\Delta \rho}{\rho})^{1/2}-1]^2$,
which are plotted in Fig.~\ref{Figrhohalpha}, as a function of $\alpha$
(we are plotting the square values, just to get positive defined
quantities, as future colliders will not know about signs for Higgs
couplings). For $\frac{\Delta\rho}{\rho}$, which depends on the parameters
of the model, we take the maximum value allowed by data~\cite{deltarho},
i.e. $\frac{\Delta\rho}{\rho}\simeq1\%$. We can appreciate that $\rho_h$
can deviate significantly from one for $\alpha\to\pi/2$, while $\rho_H$
can show large deviating from the SM prediction for $\alpha\to0$. Thus
clearly $\rho_h\neq\rho\neq\rho_H$.

Given the estimated precision expected for the measurements of the Higgs
couplings at NLC, in particular for the ratios of Higgs-Gauge couplings
which were analized in~\cite{LCp}, it happends that $\rho_{H_i}$ could be
measured with a precision of order 2 \%, which will allow to constrain
considerably the parameter $\alpha$ in this Higgs triplet model.

\bfig[t]
	\includegraphics[height=10.5cm]{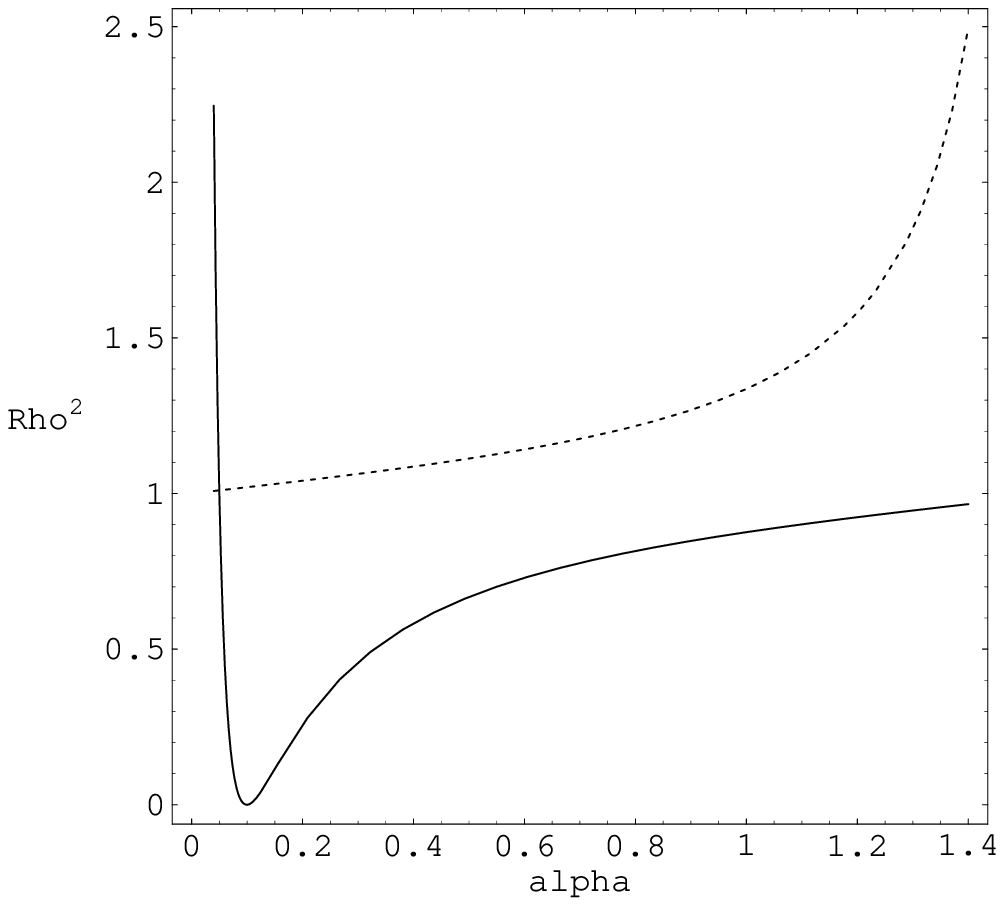}
	\caption{$\rho_h$ (dotted line) and $\rho_H$ (solid line) as a
function of the $\alpha$ mixing-angle.}
	\label{Figrhohalpha}
\efig

On the other hand, one could also use the prediction for
the Higgs decays into gauge bosons as a possible test of
violations of the custodial symmetry. For the real
decays one has:

\bea \Gamma(h\to WW)=\cos^2\alpha
        \left[1+\tan\alpha\left(\frac{\Delta
        \rho}{\rho}\right)^{1/2}\right]^2
             \Gamma(h_{SM}\to WW),\nonumber\\
\Gamma(h\to ZZ)=\cos^2\alpha\;\Gamma(h_{SM}\to ZZ),\nonumber\\
\Gamma(H\to WW)=\sin^2\alpha
        \left[1+\tan\alpha\left(\frac{\Delta
        \rho}{\rho}\right)^{1/2}\right]^2
             \Gamma(h_{SM}\to WW),\nonumber\\
\Gamma(H\to ZZ)=\sin^2\alpha\;\Gamma(h_{SM}\to ZZ).
\eea

Then the ratio $R_{\Gamma_h}=\Gamma(h\to WW)/2\Gamma(h\to ZZ)$, is
given by: \be R_{\Gamma_h}=\frac{ \Gamma(h_{SM}\to
WW)}{2\Gamma(h_{SM}\to ZZ)}
               \left[1 +\tan\alpha\left(\frac{\Delta
\rho}{\rho}\right)^{1/2}\right]^2,
\ee
while the ratio $R_{\Gamma_H}=\Gamma(H\to WW)/2\Gamma(H\to ZZ)$, is
given by: \be R_{\Gamma_H}=\frac{ \Gamma(h_{SM}\to
WW)}{2\Gamma(h_{SM}\to ZZ)}
               \left[\cot\alpha\left(\frac{\Delta
\rho}{\rho}\right)^{1/2}-1\right]^2.
\ee

This ratios are plotted in Fig.~\ref{FigRGamma} as a function of the
Higgs mass, for two fixed values of $\alpha$ (0.04 and $\pi/4$), which 
represent two typical cases of small and large mixing, respectively. In
this plot, we have included the decays into one real and one virtual gauge
boson, $(h,H)\to VV^*$, for the appropriate range of Higgs masses.

\bfig[t]
	\includegraphics[height=10.5cm]{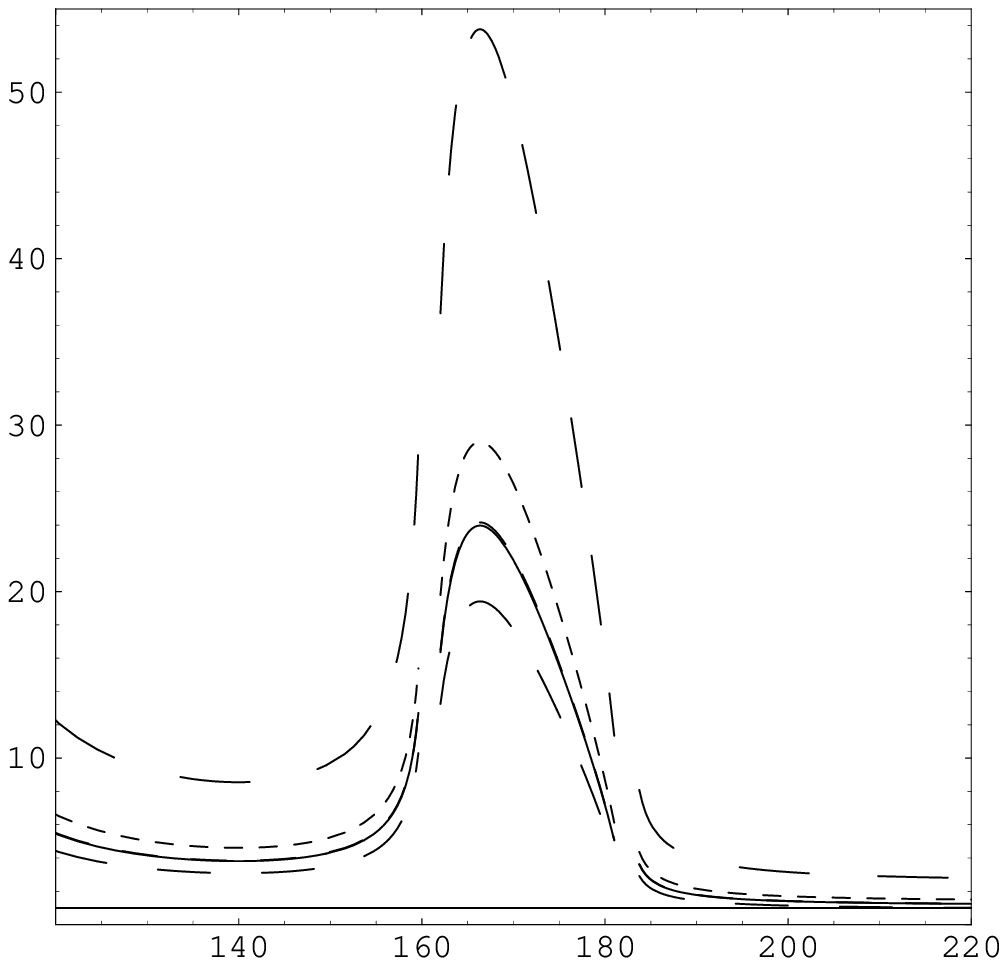}
	\caption{$R_\Gamma$ as a function of $m_h$ [GeV]. The solid line
corresponds to the SM, while the short dashed lines corresponding to
lighter Higgs ($h$). The one closer to the SM line (almost overlapping
it) corresponds to $\alpha=0.04$; the next one corresponds to 
$\alpha=\frac{\pi}{4}$. The long-dashed lines (upper and lower), 
correspond to the heavy Higgs (with mixing: $\alpha=\frac{\pi}{4}$ and
0.04, respectively). The horizontal straight line indicates the asymptotic
(SM) value.} 
	\label{FigRGamma}
\efig

\section{An extended model with custodial symmetry}
\label{sec:custodialmod} 

The Higgs sector can be extended to include extra Higgs multiplets in a
manner that respects the custodial symmetry. A minimal model with Higgs
triplets that gives $\rho=1$ was discussed in reference~\cite{georgitrip},
and studied in further detail in~\cite{GVWtrip}. This model includes a
real (Y=0) triplet, $\Xi=(\xi^+,\xi^0,\xi^-)$, and a complex (Y=2) Higgs
triplet, $\chi=(\chi^{++},\chi^+,\chi^0)$, in addition to the SM Higgs
doublet, $\Phi=(\phi^+,\phi^0)$. The v.e.v. of the neutral components can
be choossen such that $\left<\chi^0\right>=v_{T2}$, 
$\left<\xi^0\right>=v_{T0}$ and $\left<\phi^0\right>=v_D$. Then, when
$v_{T2}=v_{T0}=v_T$, the gauge boson masses are given by:
$m^2_W=m^2_Zc^2_W=\frac{1}{4}g^2 v^2$, with $v^2=v^2_D+8v^2_T$; in this
way one obtains, $\rho=1$.

The Higgs bosons can be classified according to their transformation
properties under the custodial symmetry $SU(2)_c$. The spectrum includes a
fiveplet $H^{++,+,0,-,--}_5$, a threeplet $H^{+,0,-}_3$, and two singlets
$H^0_1$ and ${H'}^0_1$. While $H^0_3$ does not couple to the gauge boson
pairs $WW$ and $ZZ$, the coupling of the remaining neutral states can be
written as:
\bea
g_{H^0_iWW}&=&gm_Wf_{H^0_i}\nonumber\\
g_{H^0_iZZ}&=&\frac{gm_W}{c^2_W}g_{H^0_i}
\eea
where the coefficients $f_{H^0_i}$ and $g_{H^0_i}$ are shown in
Table~\ref{couplingsGVW}. From this table we conclude that
$\rho_{H^0_1}=\rho_{{H'}^0_1}=1$, while $\rho_{H^0_5}=1/2$. Thus,
using our definition of the rho parameter, one can clearly
distinguish a Higgs state of the type $H^0_5$, which transforms
non-trivially under the custodial symmetry, from the states $H_1^0$
and ${H'}_1^0$, which are singlets under SU(2)$_c$. However, it should be
said that these states are not yet mass eigenstates.

\bt[b]
	\btr{||c|c|c||}\hline\hline

$H_i^0$&$f_{H_i^0}$&$g_{H_i^0}$\\\hline\hline
$H_1^0$&$c_H$&$c_H$\\\hline
$H_1^{'0}$&$\frac{2\sqrt{2}}{\sqrt{3}}s_H$&$\frac{2\sqrt{2}}{\sqrt{3}}s_H$\\\hline
$H_3^0$&0&0\\\hline
$H_5^0$&$\frac{1}{\sqrt{3}}s_H$&$-\frac{2}{\sqrt{3}}s_H$\\\hline\hline
	\etr\\
	\label{couplingsGVW}
	\caption{Coefficients $f_{H_i^0}$ and $g_{H_i^0}$ for the Higgs-Gauge
boson couplings. $t_H\equiv\frac{2\sqrt{v_{T0}^2+v_{T2}^2}}{v_D}$
.}
\et

While $H^0_1$ and ${H'}^0_1$ predict $\rho_{H_i}=1$, their couplings
with gauge bosons deviate from the SM prediction. Thus, in order
to probe this sector of the model, one could compare the decay
widths $\Gamma(H^0_i \to ZZ)$, or $\Gamma(H^0_i \to WW)$, and
using the expected precision on the Higgs measurement, determine
that range of parameters that could be excluded.

On the other hand, in terms of mass eigenstates the Higgs-Gauge boson 
couplings induce a $\rho_{H_i}$ parameter, whose expression is given by:
\be
\rho_{H_i}=\frac{\frac{1}{2}v^2_D U_{1i}+v^2_T U_{2i}+ v^2_T U_{3i} }
      {\frac{1}{2}v^2_D U_{1i}+2 v^2_T U_{2i} }
\ee
Thus, as anticipated in section 2, the choice $v_{T2}=v_{T0}=v_T$,
which makes $\rho=1$, does not imply that $\rho_{H_i}=1$. In fact, to
get $\rho_{H_i}=1$, for Higgs states that transforms as singlets under the
custodial symmetry one would need all the Higgs interactions, including
the ones appearing in the Higgs potential to respect the symmetry
SU(2)$_c$.

\section{Higgs triplets from the Little Higgs models}
\label{sec:littlehiggs}

A new approach was recently proposed to address the naturalness
problem of the Higgs sector, dubbed the ``little Higgs models'',
where the Higgs mass is protected from acquiring quadratic
divergences by being promoted as a pseudo-Goldstone boson of a
global symmetry~\cite{littleH}. The SM Higgs acquires mass via symmetry
breaking at the EW scale ($v$). While the global symmetry is broken at
high-energy scale $\Lambda_s$. The important new feature of these
models is that the Higgs remains light thanks to the global
symmetry, which includes new fields that cancel the quadratic
divergences. Furthermore, these extra Higgs fields exist as
Goldstone boson multiplets from the global symmetry breaking.

A minimal model, called the ``littlest Higgs'', is based on a
global symmetry $SU(5)$ which is broken into $SO(5)$ at the scale
$\Lambda_s=4\pi f$, while the locally gauged  subgroup is $[SU(2)\times
U(1)]^2$, which in turn breaks into the EW gauge symmetry of the SM.
This leaves 14 Goldstone bosons, including a real singlet and a real
triplet, which become the longitudinal modes of the heavy gauge
bosons, as well as a complex doublet and complex triplet, which
acquire masses radiatively, of order $v$ and $f$, respectively.
Thus, the ``littlest Higgs model'', predicts the existence of
several states with $\mathcal{O}$ (TeV) masses, which give place to
violations of the custodial symmetry~\cite{littleHph}.

Following~\cite{littleHph} one has that the light (SM-like) gauge
bosons masses contribute to the rho parameter, i.e.
$\rho=M^2_{W_L}/M^2_{Z_L}c^2_W=1+\Delta \rho$, with:
\be 
\Delta \rho=Ar^2_f+Br^2_t
\ee where
$A=\frac{5}{4}({c'}^2-{s'}^2)^2$, $B=-4$, $r_f=v/f$, and $r_t=v'/v$; $v'$
denotes the v.e.v. of the Higgs triplet of the model.

On the other hand, for the light Higgs state $h$, the model predicts
the following Higgs-Gauge couplings,
\bea\label{littleHGcouplings}
g_{hWW}&=&\frac{ig^2v}{2}\left[1+\left(\frac{1}{2}(c^2-s^2)^2-\frac{1}{3}\right)r^2_f-\frac{1}{2}s^2_0-2\sqrt{2}s_0r_t\right]\\
g_{hZZ}&=&\frac{ig^2v}{2c^2_W}\left[1-\left(\frac{5}{2}({c'}^2+{s'}^2)^2+\frac{1}{2}(c^2-s^2)^2-\frac{1}{3}\right)r^2_f-\frac{1}{2}s^2_0+4\sqrt{2}s_0r_t\right]\nonumber.
\eea
For the purpose of comparison with $\rho$, we expand $\rho_h$ in terms of
$r_f$ and $r_t$, which gives:
\be
\rho_h=1+\Delta\rho_h=1+A'r^2_f+B'r^2_t
\ee
and now: $A'= (c^2-s^2)^2+\frac{5}{2}({c'}^2-{s'}^2)^2$, and
$B'=-6\sqrt{2}s_0$. Therefore, since $A\neq A'$, $B\neq B'$ one clearly
has: $\rho\neq\rho_H$. Thus, a measurement of the Higgs couplings at NLC
will provide an independent test of the underlying symmetry of the Higgs
sector.

For instance, when $\theta=\theta'=\pi/4$ i.e. $A=0$, and $r_t=0$,
then $\Delta\rho=0$ exactly, thus, the custodial symmetry is preserved
and $\rho_h=1$ too. Furthermore, even if $r_t=r_f/4=1/20$ (maximum value
allowed in Ref.~\cite{littleHph}), one gets $\Delta\rho\simeq1\%$ which
lays within the experimental limits. In general, for values of
parameters $s_0\simeq2\sqrt{2}r_t$, $0\leq r_t<r_f/4$, $1/20\leq
r_f\leq1/5$, $1/10\leq\cot\theta=c/s\leq2$, and
$1/10\leq\tan\theta'=s'/c'\leq2$, one obtains that $\Delta\rho$ is within
the experimental limits. However, even for $\theta=\theta'=\pi/4$,
i.e. $A'=0$ and $r_t=1/20$, one gets $\rho_H\simeq0.91$ which could be
probed at NLC.

\section{Conclusions and discussion}
\label{sec:conclusion}

In this letter we proposed a definition of the rho parameter based on
the Higgs couplings with the gauge bosons, namely, 
$\rho_{H_i}\equiv\frac{g_{H_iWW}}{g_{H_iZZ}c^2_W}$, as a possible
test of the custodial symmetry. We discuss the size of such violation in
the context of general models with Higgs triplets, including the ``Little
Higgs'' model recently proposed. We find that even for Higgs models that
incorporate the custodial symmetry, to make $\rho=1$, the Higgs couplings
allow $\rho_H\neq 1$. Furthermore, in models where $\rho\simeq1$ we also
obtain that the Higgs bosons could aquire values of $\rho_H$ significantly
different from one. We find that $\rho_H$ could be tested at the level of
few percent, given the expected Higgs tests that may be achieved at the
planned next linear collider (NLC), where we will be entering into the
era of precision measurements for the Higgs sector. Violations of the
custodial symmetry could also be tested through the ratio of decay widths,
$R_{\Gamma}=\frac{\Gamma(h\to WW)}{2\Gamma(h\to ZZ)}$, with similar
precision.

In summary, given the possibility that an scalar particle could be
detected in the near future, it will be important to verify whether
this particle is indeed a type of Higgs boson, and the parameter $\rho_H$
could play a major role in this regard. This parameter meassures the
transformation properties of the Higgs bosons under the custodial
symmetry.

\begin{ack}
We akcnowledge financial support from CONACYT and SNI (M\'exico). 
Conversations with J. Erler and J. Hern\'andez-S\'anchez are
appreciated.
\end{ack}


\begin{thebibliography}{99}

\bibitem{Langacker:2002sy}
P.~Langacker,
J.\ Phys.\ G29: (2003) 1
[arXiv:hep-ph/0211065].

\bibitem{Villa:2002zt}
S.~Villa,
arXiv:hep-ph/0209359.

\bibitem{Chanowitz:2002cd}
M.~S.~Chanowitz,
Phys.\ Rev.\ D {\bf 66}, (2002) 073002
[arXiv:hep-ph/0207123].

\bibitem{Carena:2000yx}
M.~Carena {\it et al.},
arXiv:hep-ph/0010338.

\bibitem{hhunter}
J.~F.~Gunion, H.~E.~Haber, G.~L.~Kane and S.~Dawson,
{\em ``The Higgs Hunter's Guide''\/}
(Addison-Wesley Publishing Company, Redwood City, CA 1990).

\bibitem{LRmodels}
G.~Senjanovic and R.~N.~Mohapatra,
Phys.\ Rev.\ D {\bf 12}, (1975) 1502.

\bibitem{lownum}
E.~Ma and U.~Sarkar,
Phys.\ Rev.\ Lett.\ {\bf 80} (1998) 5716.
[arXiv:hep-ph/9802445].

\bibitem{littleH}
N.~Arkani-Hamed {\it et al.},
JHEP 0208 (2002) 021
[arXiv:hep-ph/0206020].

\bibitem{rhoVeltman}
M.~Veltman,
Phys.\ Lett.\ {\bf B91} (1980) 95.

\bibitem{georgitrip}
H.~Georgi and M.~Machaceck,
Nucl. Phys. {\bf B262} (1985) 463;
M.~S.~Chanowitz and M.~Golden,
Phys. Lett. {\bf B165} (1985) 105.

\bibitem{1D_11T_0}
D.~A.~L\'opez-Falc\'on, 
Rev.\ Mex.\ F\1s.\ {\bf 45} (1999) 423;
J.~L.~D\1az-Cruz and A. M\'endez,
Nucl. Phys. {\bf B380} (1992) 39.

\bibitem{deltarho}
T. Rizzo, 
Mod.\ Phys.\ Lett.\ {\bf A6} (1991) 1961.

\bibitem{LCp}
M. Bataglia and K. Desh,
[arXiv:hep-ph/0101165].

\bibitem{GVWtrip}
J.~F.~Gunion, R.~Vega and J.~Wudka,
Phys.\ Rev.\ {\bf D42} (1990) 1673.

\bibitem{littleHph}
T.~Han {\it et al.},
arXiv:hep-ph/0301040.

\end{thebibliography}
\end{document}